\documentclass[9pt,twocolumn,twoside]{osajnl}

\usepackage{url} \usepackage{hyperref} 
\usepackage[authormarkupposition=left]{changes} 

\usepackage[tight,footnotesize]{subfigure}

\journal{optica}

\setboolean{shortarticle}{false}

\usepackage{algorithm}
\usepackage{algpseudocode}
\usepackage{gensymb}

\newcommand{\stkout}[1]{\ifmmode\text{\sout{\ensuremath{#1}}}\else\sout{#1}\fi}
\setdeletedmarkup{\stkout{#1}}
\usepackage{lineno}

\title{Integrated magneto-optic based magnetometer: classical and quantum limits}

\author[1,2,3,*]{Paolo Pintus} 
\author[1,4]{Heming~Wang}
\author[5]{Sudharsanan~Srinivasan}
\author[1]{Sergio~Pinna}
\author[1,6]{Duanni~Huang}
\author[7]{Yuya~Shoji} 
\author[8]{Caroline~A.~Ross} 
\author[1]{John~E.~Bowers} 
\author[1]{Galan~Moody}

\affil[1]{Department of Electrical and Computer Engineering, University of California Santa Barbara, Santa Barbara, California 93106, USA}
\affil[2]{Department of Physics, University of Cagliari, Monserrato 09042, Italy}
\affil[3]{Central European Institute of Technology, Brno University of Technology, Brno 61200, Czech Republic}
\affil[4]{Currently with Department of Electrical Engineering and Edward L. Ginzton Laboratory, Stanford University, Stanford, California 94305, USA}
\affil[5]{Department of Electrical Engineering, Indian Institute of Technology Madras, Chennai 600 036, India}
\affil[6]{Currently with Intel Corporation, Santa Clara, California 95054, USA}
\affil[7]{Laboratory for Future Interdisciplinary Research of Science and Technology, Institute of Science Tokyo, Meguro-ku, Tokyo 152-8550, Japan}
\affil[8]{Department of Materials Science and Engineering, Massachusetts Institute of Technology, Cambridge, Massachusetts 02139, USA}
\affil[*]{Corresponding author: ppintus@ece.ucsb.edu, paolo.pintus@unica.it}

\begin{abstract}
Magnetic field sensors with high sensitivity and spatial resolution have profoundly impacted diverse applications ranging from geo-positioning and navigation to medical imaging, materials science, and space exploration. However, the use of high-precision magnetometers is often limited due to their bulky size or low energy efficiency. In this work, we present the design, modeling and an experimental demonstration of an all-optical magnetometer based on silicon integrated photonics heterogeneously integrated with a magneto-optic thin film. By bonding a thin cerium-yttrium iron garnet layer onto an integrated silicon photonic interferometer, small magnetic field fluctuations can be detected through the non-reciprocal phase shift in the sensor. This strategy enables more than 80 dB of dynamic range with better than 40~pT/$\sqrt{\text{Hz}}$ sensitivity at room temperature. Importantly, by leveraging silicon photonics, the core platform is scalable through foundry manufacturing, and the ultra-low power requirements enable complete system integration with on-chip lasers, detectors, and quantum elements for enhanced sensitivity. This work provides a path to realizing a compact, scalable, room temperature magnetometer based on integrated photonic systems, opening new opportunities for ultra-sensitive and ultra-efficient magnetic field detectors.
\end{abstract}

\setboolean{displaycopyright}{true}

\begin{document}
\maketitle

\section{Introduction}

Magnetometers enable measurements of local variations in magnetic field strength and orientation, serving as important sensors for many applications including geopositioning and navigation~\cite{canciani2016absolute,claussen2020magnetic}, biomedical research~\cite{kitching2018chip}, and space exploration~\cite{bennett2021precision}. Although the use of magnetometers for navigation dates back to the 11th century when compasses were first employed, modern sensors rely on more sophisticated techniques and instrumentation, such as superconducting quantum interference devices (SQUIDs)~\cite{fagaly2006superconducting}, atomic vapor cells (AVCs)~\cite{robbes2006highly}, nitrogen-vacancy centers (NV)~\cite{rondin2014magnetometry}, and spin-exchange relaxation-free (SERF) magnetometers~\cite{allred2002high}. Although high-precision magnetometers are desirable, their use is often limited by the large size, weight, and power (SWaP) of the sensors and auxiliary systems. SQUIDs need to be operated in a cryogenic environment and are thus incompatible with small-SWaP applications. High-sensitivity NV magnetometers require bulky instrumentation, which currently limits their deployment outside laboratory environments, while portable diamond sensors still exhibit relatively large noise floors, ranging from tens of $\mu$T/$\sqrt{\text{Hz}}$~\cite{kim2019cmos} 
down to hundreds of pT/$\sqrt{\text{Hz}}$~\cite{sturner2021integrated}. Meanwhile, miniaturized AVC and SERF magnetometer systems are still limited by their mass (hundred grams) and high-power requirements  (1~W to several watts), which hamper their utilization in portable applications~\cite{korth2016miniature,oelsner2022integrated}.

Sensors based on photonic integrated circuits (PICs) offer reduced SWaP and can outperform conventional magnetometer platforms by enabling superior integration, room-temperature operation, and seamless interfacing with compact optical components and subsystems~\cite{thomson2016roadmap,sturner2021integrated,zhou2023prospects}. Furthermore, an all-optical chip-scale magnetometer could simultaneously lead to high sensitivity and straightforward integration with quantum elements such as squeezed light sources and quantum emitters~\cite{moody20222022,moody2020chip}. Thus, advancing chip-scale magnetometers towards room-temperature operation with low-power requirements could pave the way for new frontiers of scientific research and cutting-edge applications.

Here, we present the design, simulation and experimental verification of an all-optical magnetometer based on a chip-scale photonic interferometer combined with a heterogeneously integrated magneto-optic (MO) thin film (see Fig. \ref{device_layout} for a schematic illustration). At optical frequencies, the permittivity tensor of these materials changes in response to an external magnetic field, subsequently influencing the polarization, attenuation, and phase of light as it propagates through the PIC. Focusing on the nonreciprocal phase shift (NRPS) effect, our sensor utilizes an integrated interferometer to accurately determine both the direction and intensity of an external magnetic field. Our approach combines a silicon photonic Mach-Zehnder interferometer (MZI) with a bonded cerium-yttrium iron garnet (Ce:YIG) thin film, which reduces the SWaP and cost of magnetic sensors by leveraging light-based technology to offer high speed, energy efficiency, high resolution, room-temperature operation, and negligible electromagnetic noise. We demonstrate a dynamic range of 80~dB with a sensitivity of 40~pT/$\sqrt{\text{Hz}}$, limited primarily by thermo-refractive noise. This remarkable level of sensitivity, dynamic range, and compact form factor, together with operation at room temperature, could mark a significant transformation in the field of high-precision magnetometry.

In what follows, we provide the theoretical framework for PIC-based MO sensing, including the working principle, noise contributions, and device design. We then discuss the fabrication and characterization of the MO-PIC, including linear optical characterization to measure and distinguish between magnetic and thermal signal contributions. We conclude with a discussion on the achievable sensitivity and fundamental limiting factors (thermo-refractive and shot noise), including prospects for further improvement through the injection of squeezed light into the MO-PIC, analogous to the approach utilized in advanced LIGO for gravitational wave detection ~\cite{ganapathy2023broadband,aasi2015advanced}.

\section{Working principle and device layout}

When an optical wave travels through an MO material, it can undergo alterations due to the presence of an external magnetic field. Depending on the direction of propagation and the orientation of the magnetic field, the light can experience various nonreciprocal effects, such as the rotation of the polarization plane, variation of the phase velocity, or changes in the propagation loss~\cite{dotsch2005applications,landau2013electrodynamics}. 

The proposed integrated photonic magnetometer capitalizes on the nonreciprocal phase shift (NRPS) effect observed when the magnetic field is perpendicular to the direction of light propagation (Voigt configuration). This setup results in varying phase velocities depending on both the direction of light propagation (forward and backward) and the orientation of the magnetic field. In the past, this effect has been utilized to create integrated optical isolators and circulators~\cite{tien2011silicon,bi2011chip,huang2016electrically,pintus2017microring,huang2017dynamically,zhang2019monolithic,pintus2019broadband}, magneto-optic modulators and switches~\cite{pintus2022integrated,rombouts2023sub,yajima2023high}, as well as optical memories~\cite{murai2020nonvolatile,pintus2025integrated}. Here, we demonstrate how NRPS can be effectively employed to detect external magnetic fields.

\begin{figure}[t!]
\centering\includegraphics[width=1.0\columnwidth]{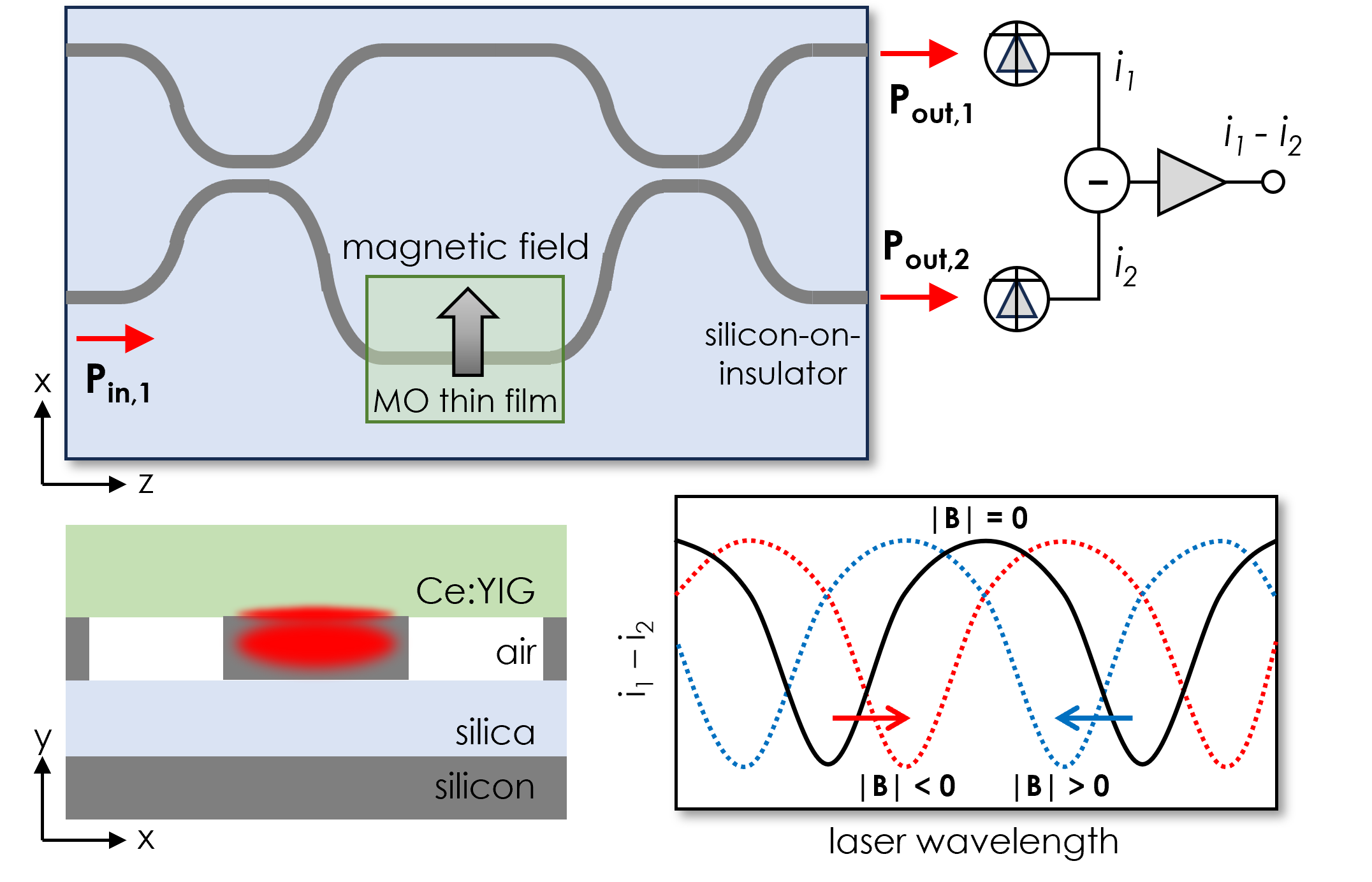}
\caption{\label{device_layout} Concept for photonic-integrated magneto-optic sensor. The top panel illustrates the device schematic. Laser light ($P_{in,1}$) into the asymmetric Mach-Zehnder interferometer (MZI) propagates along the two arms, with a magneto-optic (MO) thin film bonded above one arm. Magnetic field fluctuations impart a nonreciprocal phase shift, which is electronically detected using a balanced homodyne detector. The bottom-left panel depicts the cross-section of the structure with the MO material (Ce:YIG). The optical mode is indicated in red, which partially overlaps with the MO material. The bottom-right panel illustrates the detected response from the MZI versus in-plane magnetic field orientation.}
\end{figure}

The proposed magnetometer is shown schematically in Fig.~\ref{device_layout}. An MO material is placed on one of the arms of an unbalanced Mach-Zehnder Interferometer (MZI). By varying the wavelength of the input laser, characteristic interference fringes are observed in the output power at each photodetector (PD). When a magnetic field is switched on, the light propagating in the arm with the MO material is influenced by the NRPS effect, which depends on the amplitude and direction of the external magnetic field. Consequently, the power at the photodetector changes compared to the case of no magnetic field. By measuring the spectrum shift or the power drop, we can detect the amplitude and direction of the external magnetic field.

The phase delay and amplitude attenuation depend on the optical length and propagation loss along the two arms of the MZI. Let us assume that arm 1 (arm 2) has a length of $L_1$ ($L_2$), while the corresponding phase constant and field attenuation are $\beta_1$ ($\beta_2$) and $\alpha_1$ ($\alpha_2$), respectively. In the device under investigation, the two directional couplers are designed to be identical and light is equally split and recombined between the two arms. If the light is launched from only one input port, the photocurrent at the PDs placed at the MZI outputs is
\begin{subequations}\label{eq:1}
\begin{align}
I_{1} & = \dfrac{\eta R_{\lambda}A\, P_{in,1}}{2}\left[ 
\cosh{\left(\delta_\alpha\right)}
-\cos{\left(\delta_\beta\right)}\right] \\ 
I_{2} & = \dfrac{\eta R_{\lambda}A\, P_{in,1}}{2}\left[ 
\cosh{\left(\delta_\alpha\right)} + 
\cos{\left(\delta_\beta\right)}\right],
\end{align}
\end{subequations}
where $\eta$ is the coupling efficiency, $R_\lambda$ represents the responsivity of the PDs, $A$ is the average power attenuation in the MZI, $\delta_\alpha$ denotes the difference in field attenuation between the two arms, and $\delta_\beta$ stands for the optical length difference between the two arms. More formally, the terms $A$, $\delta_\alpha$, and $\delta_\beta$ are defined as
\begin{subequations}\label{eq:2}
\begin{align}
A &=\exp{\left[-\left(\alpha_{1}L_1+\alpha_{2}L_2\right)\right]}\label{attenuation}\\
\delta_\alpha & = \alpha_{1}L_1-\alpha_{2}L_2\label{delta_loss}\\
\delta_\beta  & = \beta_{1}L_1-\beta_{2}L_2\label{delta_beta}.
\end{align}
\end{subequations}

Having two identical directional couplers with a power splitting ratio $K=0.5$ is crucial to eliminate common noise, such as the relative intensity noise of the laser. However, due to fabrication variations, the actual splitting ratio deviates from the nominal value. Defining the perturbation of the coupling power as $\Delta K$ to account for unequal splitting, \eqref{eq:1} is modified as follows
\begin{subequations}\label{eq:1bis}
\begin{align}
I_{1}  = \dfrac{\eta R_\lambda A\,P_{in,1}}{2}&\left[
\left(1 + 4\Delta K^2\right)\cosh{\left(\delta_\alpha\right)}
+4\Delta K\sinh{\left(\delta_\alpha\right)}\right.\nonumber \\
&\left.-\left(1 - 4\Delta K^2\right)\cos{\left(\delta_\beta\right)}\right]\\ 
I_{2} = \dfrac{\eta R_\lambda A\,P_{in,1}}{2}&\left[  
\left(1 - 4\Delta K^2\right) \cosh{\left(\delta_\alpha\right)}\right.\nonumber \\
&\left.+\left(1 - 4\Delta K^2\right)\cos{\left(\delta_\beta\right)}\right], 
\end{align}
\end{subequations}
where $\Delta K$ can assume positive and negative values around zero, i.e.,  no error. For the sake of clarity, derivations of \eqref{eq:1} and \eqref{eq:1bis} are reported in the supplementary material.

The proposed magnetometer exploits the use of a balanced PD which can effectively minimize the impact of common noise that affects both output ports of the MZI. In the suggested configuration, the resulting output current of the balanced PD is
\begin{align}\label{eq:3}
I_{out}= \left[I_{1}-I_{2}\right]=
&\eta R_\lambda A\, P_{in,1}\left[2\Delta K\sinh{\left(\delta_\alpha\right)}
+4\Delta K^2\cosh{\left(\delta_\alpha\right)}\right]\nonumber\\
&-\eta R_\lambda A\, P_{in,1}\left(1 - 4\Delta K^2\right)\cos{\left(\delta_\beta\right)}.
\end{align}


To attain the maximum sensitivity of the magnetometer, let us consider the ideal scenario where $\Delta K =0$. Under this condition, the highest sensitivity is achieved by biasing the argument of the cosine terms in \eqref{eq:3} by $\pi/2$. This can be obtained either by designing one arm of the MZI to be longer or by incorporating an integrated phase shifter on the arm where no magnetic material is located. As such, the optical length difference between the two arms is given by 
\begin{equation}\label{eq:90deg_offset}
\delta_\beta=\dfrac{\pi}{2}+\Delta\phi(H), 
\end{equation}
where $\Delta\phi(H)$ is the variation of the optical length introduced by the magnetic field. 

The fundamental factor that determines the quality of a magnetometer lies in the conversion gain, $G$. This gain establishes the connection between changes in the magnetic field and the resultant detected signal, the current at the PD or the phase in the MO arm. Within our specific context, we define $G$ as the ratio of NRPS to the variation in magnetic field intensity per unit length:
\begin{equation}\label{eq:gain}
G=\dfrac{1}{L}\dfrac{\partial\phi}{\partial H}.
\end{equation}
Combining \eqref{eq:90deg_offset} and \eqref{eq:gain} in \eqref{eq:3}, the current measured at the balanced PD produced by a small field is
\begin{align}\label{eq:Iout}
I_{out}&=\eta R_\lambda A\, P_{in,1}\sin{\left[\Delta\phi(H)\right]}\approx \eta R_\lambda A\, P_{in,1} G\,L\,H,
\end{align}
where $L$ is the length of the MO layer. 

\section{Noise contributions}

As shown in the previous section, the intensity of the magnetic field can be detected by measuring the current of the balanced photodetector. Nevertheless, phase fluctuation due to temperature, the power fluctuation of the input laser, and photodetector noise can limit the smallest magnetic field that can be detected accurately.

The impact of those noise sources at the two PDs can be determine by differentiating \eqref{eq:1bis}:
\begin{subequations}\label{eq:7}
\begin{align}
\Delta I_{1}  = & +\dfrac{\eta R_{\lambda}A\,P_{in,1}}{2}\left(1 - 4\Delta K^2\right)\Delta\phi \\
&+ \dfrac{\eta R_{\lambda}A}{2}\left[
\left(1 + 4\Delta K^2\right)\cosh{\left(\delta_\alpha\right)}
+4\Delta K\sinh{\left(\delta_\alpha\right)}\right]\Delta P_{in,1}\label{common_mode_DC_I1}\\
& +\dfrac{\eta R_{\lambda}A}{2}\left(1 - 4\Delta K^2\right)\Delta\phi \Delta P_{in,1}\label{common_mode_AC_I1}\\
& + \Delta I_{noise,1} \label{eq7:P1}
\end{align}
\end{subequations}
and
\begin{subequations}\label{eq:8}
\begin{align}
\Delta I_{2}  = & - \dfrac{\eta R_{\lambda}A\,P_{in,1}}{2}\left(1 - 4\Delta K^2\right)\Delta\phi \\
& + \dfrac{\eta R_{\lambda}A}{2}
\left(1 - 4\Delta K^2\right) \cosh{\left(\delta_\alpha\right)}\Delta P_{in,1} \label{common_mode_DC_I2}\\
& - \dfrac{\eta R_{\lambda}A}{2}\left(1 - 4\Delta K^2\right)\Delta\phi\Delta P_{in,1}\label{common_mode_AC_I2}\\
& + \Delta I_{noise,2}. \label{eq7:P2}
\end{align}
\end{subequations}

The phase noise, $\Delta\phi$, is introduced by temperature fluctuations in two distinct ways: directly through the {\it thermo-refractive noise}, and indirectly via the {\it thermal fluctuations of the magnetic domains}. The main source of common noise arises from the {\it relative intensity noise} (RIN) of the laser coupled to the interferometer. This contributing term is represented by $\Delta P_{in,1}$ in equations \eqref{eq:7} and \eqref{eq:8}. Finally, $\Delta I_{noise,1}$ and $\Delta I_{noise,2}$  pertain to noise within the photodetectors, which can arise in two forms: {\it shot noise} and {\it thermal noise}. The specific details of each noise contribution are elaborated in the subsequent discussion.

\subsection{Thermo-refractive noise}

Thermodynamic fluctuation of temperature along a fiber or a waveguide produces a local variation of the refractive index, resulting in a time-variant phase delay~\cite{wanser1992fundamental, bartolo2012thermal}. The autocorrelation function for the phase due to temperature fluctuation is written as 
\begin{equation}
\langle\phi(t_1),\phi(t_2)\rangle = \int_{-\infty}^{\infty}\text{SD}^{T}_{\phi\phi}(\omega)\,e^{i\omega (t_1-t_2)}\,d\omega,
\end{equation}
where $\text{SD}^{T}_{\phi\phi}(\omega)$ is the power spectral density function associated with phase variation $\Delta\phi(t)$ (Wiener-Khinchin theorem). 

In order to calculate the power spectral density of phase fluctuations, we utilize the fluctuation-dissipation theorem. This involves treating temperature fluctuations as the {\it observable output} of the system and entropy as the corresponding {\it conjugate input force}~\cite{callen1951irreversibility, de2013non, levin2008fluctuation,landau2013statistical}. Our approach involves solving the heat equation for our system while incorporating a fictitious entropy that acts as an external stimulus. The power spectral density is given by
\begin{equation}\label{eq:SD_thermo-refractive}
    \text{SD}^T_{\phi\phi}(\omega) = \dfrac{4k_B T}{\omega^2}\left(\dfrac{W_{\text{diss}}}{A_s^2}\right)\left(\dfrac{\omega_o n_{g}L}{cN}\right)^2
\end{equation}
where $k_B$ denotes the Boltzmann constant, $T$ is the system temperature, $\omega$ represents the angular frequency of the noise, and $c$ stands for the speed of light. The first and second terms enclosed within the brackets have dependencies on the waveguide thermal and optical properties, respectively. The first term is determined through numerical solutions of the heat equation. In this context, $A_s$ represents the amplitude of the input entropy, and $W_{\text{diss}}$ corresponds to the power dissipated during the heat transfer. Regarding the second part of the expression, $\omega_o$ refers to the optical angular frequency, $n_{g}$ is the group index, and $N$ serves as a normalization factor for the optical mode.
Derivation of \eqref{eq:SD_thermo-refractive} and additional details regarding the modeling and calculation are provided in section~4 of the supplementary material.

The resulting currents at the PDs are
\begin{equation}\label{eq:current_SD_TR}
    \Delta I_{i,tr}=(-1)^{i+1}\dfrac{\eta R_{\lambda}A\,P_{in,1}}{2}\left(1 - 4\Delta K^2\right)\sqrt{4\pi\,\text{SD}^T_{\phi\phi}(0)\,B}
\end{equation}
where $B$ is the single-sided bandwidth of the detector (in Hz) over which the noise is considered. The factor $4\pi$ within the square root addresses the conversion of $B$ from Hz to rad/s and accounts the integration of the two-sided spectrum. Here, we assumed $B$ is smaller than the dynamic range of the noise. 

\begin{figure*}[thb!]
\centering
\subfigure[\protect   \label{Mode_Profile}]{
\includegraphics[width=0.45\textwidth]{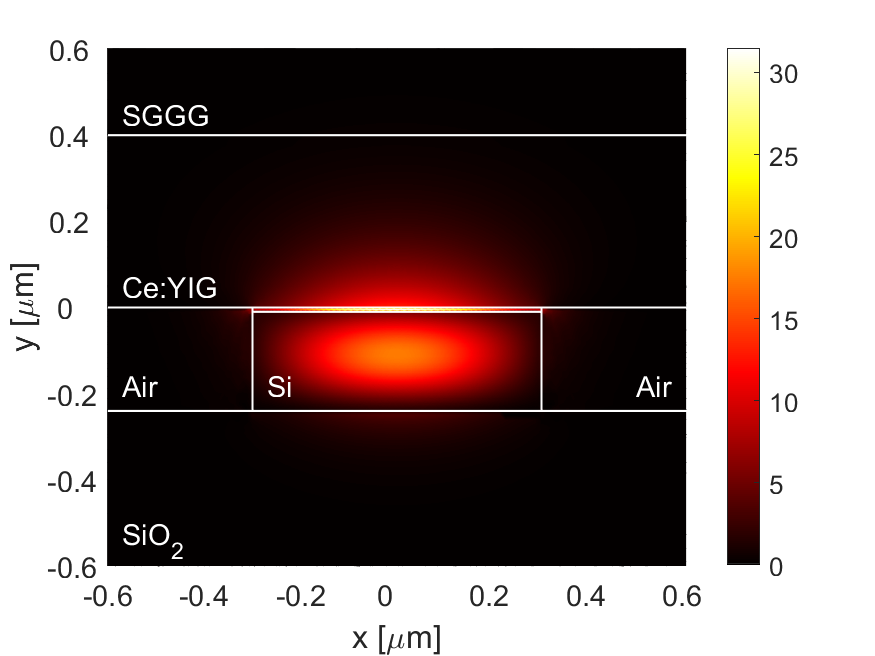}}
\subfigure[\protect  \label{NRPS_PAPER}]{
\includegraphics[width=0.45\textwidth]{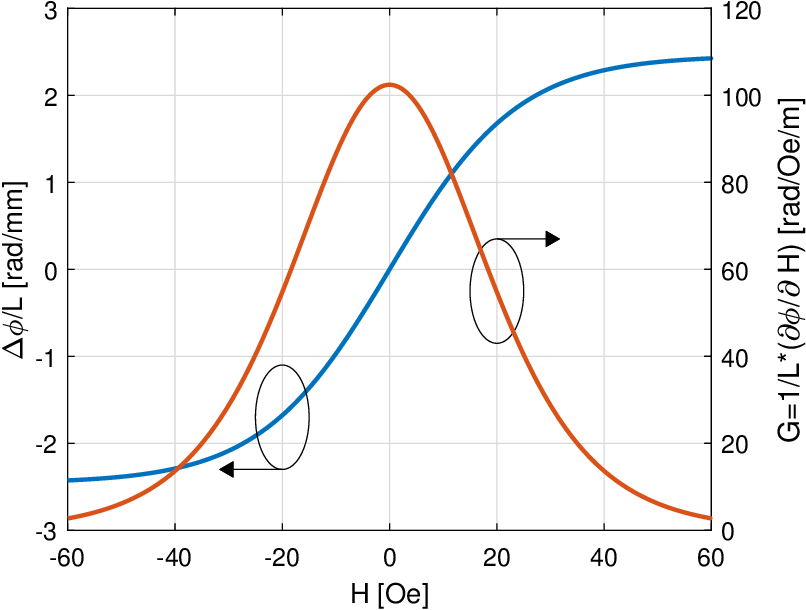}}
\\
\subfigure[\protect  \label{PhaseNoiseSpectrum}]{
\includegraphics[width=0.45\textwidth]{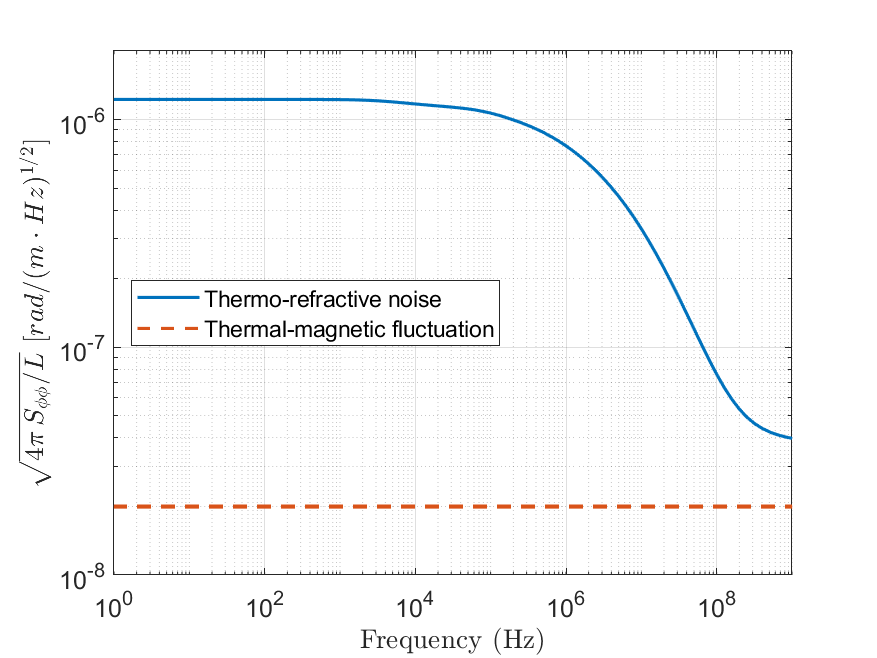}}
\subfigure[\protect  \label{Sensitivity_Limit_PAPER}]{
\includegraphics[width=0.45\textwidth]{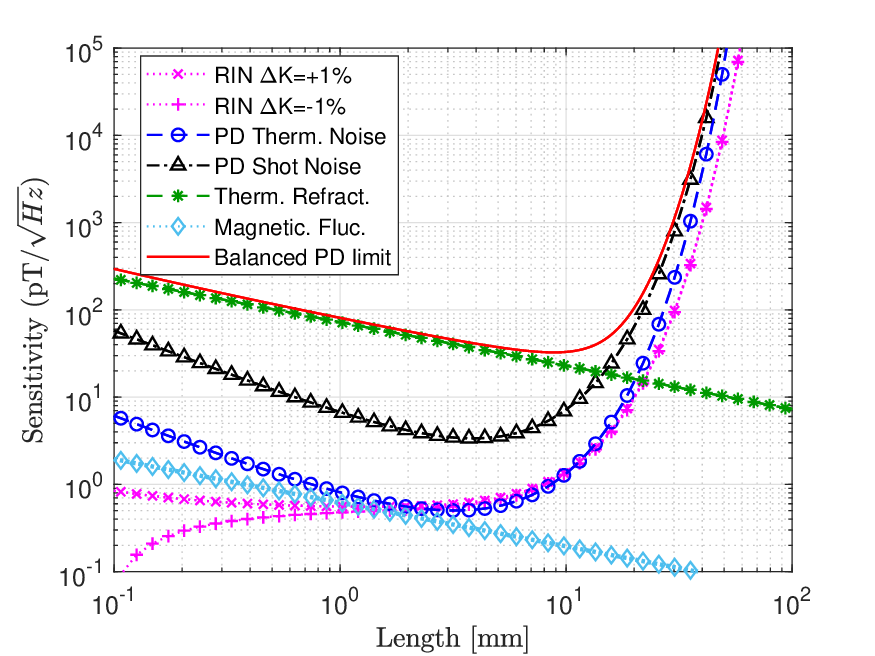}}
\caption{Simulated device performance for heterogeneous Ce:YIG-on-Si integrated magnetometer. (a)~Transverse magnetic (TM) mode profile, where a silica layer of 10 nm is assumed between the silicon waveguide and the bonded Ce:YIG layer. (b)~Phase change per unit length (left-axis) and non-reciprocal phase shift per unit magnetic field intensity and per unit length (right-axis). (c)~Root mean square of the power spectral density per unit length of the phase variation due to the thermo-refractive noise and thermal-magnetic fluctuation. (d)~Simulated contributions from various noise sources
to the sensitivity as a function of sensor length for $P_{in,1}$ = 20~mW}\label{Fig:Sim_CeYIH_on_Si}
\end{figure*}

\subsection{Thermal-magnetic fluctuation}

Thermal fluctuations can also influence the magnetization of the MO material. Given that magnetization governs the nonreciprocal phase shift, these fluctuations will result in phase variations.

The thermal-magnetic fluctuation can be represented by an effective magnetic field that maintains a zero-time average, exhibits temporal and spatial decorrelation, and oscillates within a volume $V$, i.e., the domain where thermal fluctuations take effect~\cite{brown1963thermal, leliaert2017adaptively}. The associated power spectral density is written as
\begin{equation}\label{eq:SD_magnetic-fluctuation}
\text{SD}^M_{\phi\phi}(\omega)=2\kappa_B T\, G^2L_{cell}^2\left(\dfrac{1}{\mu_{ii}\mu_0}\sqrt{\dfrac{\alpha_{G}}{M_s\gamma\,V}}\right)^2 \left(\dfrac{A_{\text{eff}}\,L}{V}\right)
\end{equation}
where $\gamma$ is the gyromagnetic ratio, $\alpha_G$ is the dimensionless Gilbert damping parameter, $M_s$ the saturation magnetization, $L_{cell}$ and $V$ are the size and volume of the single magnetic domain in the MO, respectively. Additionally, $\mu_0$ corresponds to the magnetic permeability of vacuum, while $\mu_{ii}$ signifies the relative permeability of the MO material. Here, the index $i$ spans the Cartesian coordinate system $x$, $y$, and $z$ axes and it depends on the direction of the magnetic field considered for the NRPS. Similar to \eqref{eq:SD_thermo-refractive}, the first term within the brackets is contingent upon the magnetic characteristics of the material. Meanwhile, the second term concerns the overlap between the optical mode and the volume of each thermal-magnetic fluctuation. Here, $A_{\text{eff}}$ represents the effective area where the mode intersects with the MO material. If $L_{ex}$ is the magnetic exchange length~\cite{abo2013definition}, we assume $L_{cell}=2\,L_{ex}$ and $V=L_{cell}^3$. The derivation of \eqref{eq:SD_magnetic-fluctuation} is reported in section~5 of the supplementary material.

Similar to the previous scenario, the noise currents at the PDs produced by the thermal-magnetic fluctuation can be written as
\begin{equation}\label{eq:current_SD_MF}
    \Delta I_{i,mf}=(-1)^{i+1}\dfrac{\eta R_{\lambda}A\,P_{in,1}}{2}\left(1 - 4\Delta K^2\right)\sqrt{4\pi\, \text{SD}^M_{\phi\phi}(0)\,B}
\end{equation}

\subsection{Laser relative intensity noise}

Variations in laser intensity can also limit the smallest detectable magnetic field. The impact of laser RIN on integrated optical sensors has been previously studied~\cite{srinivasan2014integrated,srinivasan2014design,huang2015compact,gundavarapu2017interferometric}. In the current magnetometer configuration, we implement balanced photodetection to minimize common input fluctuations. From \eqref{eq:7} and \eqref{eq:8}, the contribution of the RIN=$\Delta P_{in,1}/P_{in,1}$ to the output current of the balanced PD is
\begin{align}
\Delta I_{out}=&  \eta R_{\lambda}A \,P_{in,1}\left[
4\Delta K^2\cosh{\left(\delta_\alpha\right)}
+2\Delta K\sinh{\left(\delta_\alpha\right)}\right]\text{RIN}\nonumber \\
&+\eta R_{\lambda}A\,P_{in,1}\left(1 - 4\Delta K^2\right)\Delta\phi\,\text{RIN},
\end{align}
which can affect both the common mode, \eqref{common_mode_DC_I1} and \eqref{common_mode_DC_I2}, and the signal, \eqref{common_mode_AC_I1} and \eqref{common_mode_AC_I2}. 
However, the impact on the terms \eqref{common_mode_AC_I1} and \eqref{common_mode_AC_I2} can be neglected, as their contribution is of higher order and negligible. As observed, when $\Delta K = 0$, the common mode cancels out and the RIN does not affect the measured current.

\subsection{PD thermal noise}
Thermal noise can also directly impact the PDs due to thermal fluctuations occurring within electron density. In the prevalent formulation, this noise is modeled with a voltage generator inducing fluctuations and connected at the terminals of a noiseless resistor (Thevenin’s circuit). The mean squared value of these voltage fluctuations is represented as $\Delta V = \sqrt{4k_BT\,R\,B}$, where $R$ is the resistance of the PD. Converting $\Delta V$ in the root mean square of an equivalent current generator (Norton's circuit), we obtain 
\begin{equation}
    \Delta I_{i,th}= \sqrt{\dfrac{4\kappa_B TB}{R}}\quad\text{with}\quad i=1,2
\end{equation}

\subsection{PD Shot noise}
Shot noise arises in situations involving potential barriers, such as within a pn-junction of a PD, and stems from the discrete nature of electron charge. While the impact of other noise sources can be notably reduced by lowering the temperature, shot noise stands as an intrinsic quantum limitation of detectors, limiting the minimum magnetic fields that can be detected using classical light fields in the MZI. The root mean squared value of the shot noise current is
\begin{equation}
    \Delta I_{i,sh}= \sqrt{2\, q\,|I_i|\,B},\quad\text{with}\quad i=1,2
\end{equation}
where $q$ is the elementary charge of an electron, and $I_i$ is the average DC current flowing the PDs, as defined in \eqref{eq:1bis} when $\delta_{\beta}=\pi/2$. 

\subsection{Magnetic field sensitivity}

The sensitivity represents the minimum detectable magnetic field and serves as a pivotal aspect of any magnetometer. In our optical magnetometer, the lower limit of the measurable magnetic field hinges on the noise level of the PD current. When no magnetic field is applied, the current of the two PDs can be assumed identical and approximately equal to $R_\lambda A P_{in,1}/2$. Using \eqref{eq:Iout}, the minimum detectable magnetic field is
\begin{equation}\label{eq:sensitivity}
\begin{aligned}
    H_{m}\approx\dfrac{\sqrt{B}}{GL} &\left[\sqrt{4\pi\,\text{SD}^T_{\phi\phi}} + \sqrt{4\pi\,\text{SD}^M_{\phi\phi}} \right. \\
    & \left.+\dfrac{\sqrt{8\kappa_B T}}{R_\lambda A P_{in,1}\sqrt{R}}
    + \sqrt{\dfrac{16q}{R_\lambda A P_{in,1}}}\right]\\
+\dfrac{1}{GL}&\left[
4\Delta K^2\cosh{\left(\delta_\alpha\right)}
+2\Delta K\sinh{\left(\delta_\alpha\right)}\right]\text{RIN}
\end{aligned}
\end{equation}
where we make the assumption $\Delta K\simeq0$ for the sake of simplicity. In the previous formula, we also assume that the shot noise and thermal noise currents of the PD are two distinct and independent random variables. As a result, the variance of their difference is equal to the sum of their individual variances.

\section{Modelling of heterogeneous Ce:YIG-on-Si magnetometer}

To assess the performance of the proposed optical magnetometer, we utilize a silicon-on-insulator (SOI) platform for the fabrication of the MZI with a cerium-substituted yttrium iron garnet (Ce:YIG) bonded on top. Ce:YIG is a commonly used MO material that is optically transparent to telecom wavelengths ($\sim$1550~nm) and has one of the largest Faraday rotation constants in this wavelength range, $\theta_F=4500\degree$cm$^{-1}$. Using a material with large $\theta_F$ is highly important, as the conversion gain $G$ is directly proportional to it.


In a silicon waveguide with a bonded Ce:YIG layer, the fundamental transverse magnetic (TM) mode exhibits a NRPS when the external magnetic field lies in the plane of the MO material and is perpendicular to the direction of light propagation. However, no NRPS is discernible when the external magnetic field is oriented differently. As for the fundamental transverse electric (TE) mode, the symmetry of the waveguide cross-section prevents any noteworthy NRPS from occurring~\cite{pintus2013integrated}. To maximize this effect, the cross-section of the waveguide has been optimized, resulting in a 600~nm wide and 220~nm tall silicon waveguide with a Ce:YIG layer thicker than 400~nm~\cite{pintus2011design,pintus2014accurate}. In our computation, we have also considered an SiO$_2$ layer between Ce:YIG and Si. This layer is a byproduct of the bonding process and its thickness can significantly reduce the NRPS~\cite{pintus2017microring}. With a silica boding layer as small as 10~nm, the mode profile of the TM fundamental mode in the optimal case is shown in Fig.~\ref{Mode_Profile}. 

The central feature of this class of magnetometers is the variation of phase when an external magnetic field is applied. Fig.~\ref{NRPS_PAPER} shows the NRPS per unit length of the TM mode when a magnetic field directed along the $x$-axis is applied [refer to Fig.~\ref{Mode_Profile}]. The phase change demonstrates a nearly linear behavior within the range of $\pm$15~Oe ($\pm$1.5~mT), while saturation occurs for magnetic fields larger than 40~Oe (4~mT). From the same simulation, the computation of the conversion gain is straightforward. When the magnetic field approaches zero, the gain reaches its peak, exceeding 100~rad/Oe/m (1~rad/mT/mm).

To evaluate the phase noise of the proposed magnetometer, we calculated the power spectral density of the thermo-refractive noise and thermal-magnetic fluctuation for the TM mode, which is shown in Fig.~\ref{PhaseNoiseSpectrum} (see supplementary material for more details). As shown in the figure, the thermo-refractive noise dominates the thermal-magnetic fluctuation. We also observed that the root mean squared value of the thermo-refractive noise per unit length is approximately 20 times greater than the one measured in the fiber at the same wavelength \cite{wanser1992fundamental}. This difference aligns well with the substantial variation in thermo-optic coefficients between silicon ($dn/dT=$1.86$\times 10^{-4}$~K$^{-1}$, \cite{komma2012thermo}) and silica ($dn/dT=$1.0$\times 10^{-5}$~K$^{-1}$, \cite{elshaari2016thermo}). Operating at lower temperature can effectively reduce such noise contributions. 

We evaluated the maximum sensitivity with respect to the length of the arms of the MZI using \eqref{eq:sensitivity}. As depicted in Fig.~\ref{Sensitivity_Limit_PAPER}, we compute the contributions to the minimum detectable magnetic field. For this evaluation, we assume that the two arms of the MZI have a similar length, and the propagation loss for the TM mode in the Ce:YIG-on-Si waveguide is 25~dB/cm~\cite{pintus2022integrated}. All the parameters used for the simulations are shown in Table~\ref{tab:sim_params}. In this scenario, for a device shorter than 1~cm, we can reach the minimum spectral sensitivity of 32~pT$\cdot$Hz$^{-1/2}$. By integrating over one second (i.e., assuming a bandwidth of 1~Hz), this corresponds to a magnetic field sensitivity of 32~pT.

\begin{table}[htb]
\centering
\begin{tabular}{|l||c|}
\hline
\textbf{Parameter} & \textbf{Value}\\
\hline\hline
Input laser RIN                         & -140~dB/Hz  \\
Input optical power $P_0$               & 20~mW       \\
Loss MO arm $\alpha_1$                  & 25.0~dB/cm  \\
Loss passive arm $\alpha_2$             & 2.0~dB/cm   \\
MZI arm imbalance $L_1-L_2$             & 143~$\mu$m  \\
Perturbation coupling coeff. $\Delta K$ & $\pm$0.01   \\
Coupling efficiency $\eta$              & 1           \\
PD responsivity $R_{\lambda}$ (InGaAs)  & 0.95~A/W    \\
PD resistance $R$                       & 1~k$\Omega$ \\
PD bandwidth $B$                        & 1~Hz        \\
\hline
\end{tabular}
\caption{Simulation parameters used for the MZI magnetometer model.}
\label{tab:sim_params}
\end{table}

The simulated noise contributions shown in Fig.\ref{Sensitivity_Limit_PAPER} were calculated for the configuration in Fig.\ref{device_layout}, where the MO material is included only in one arm of the MZI. This asymmetry breaks the loss balance between the two arms, increasing the contribution of the RIN. However, when $\Delta K \leq 0.01$, this contribution remains below the shot noise and thermal noise of the PD. We verified that in case of balanced losses (e.g., $\alpha_1 = \alpha_2 = 25$dB/cm), the RIN contribution decreases significantly, and the sensitivity remains nearly unchanged (about 38pT$\cdot$Hz$^{-1/2}$).

Figure \ref{Sensitivity_Limit_PAPER} and \eqref{eq:sensitivity} play a pivotal role in comprehending and improving the proposed magnetometer. Here, we emphasize several crucial aspects: (i)~the contribution to phase noise can be effectively mitigated by extending the length of the MO material without an upper limit; (ii)~the thermal and shot noise of the PDs exhibit an optimal length. Extending the arm length leads to a greater phase shift [\eqref{eq:Iout}], but simultaneously, it induces pronounced signal attenuation, [$A$ defined in \eqref{attenuation}]; (iii)~the most substantial constraint of the present platform originates from thermo-refractive noise, predominantly attributed to the substantial thermo-optic coefficient of silicon at room temperature; (iv)~larger conversion gain, $G$, can significantly improve the sensitivity of this class of optical magnetometers.

\section{Experimental demonstration}

\begin{figure*}[!thb]
\centering
\subfigure[\protect   \label{Fabricated_MZIs}]{
\includegraphics[width=0.41\textwidth]{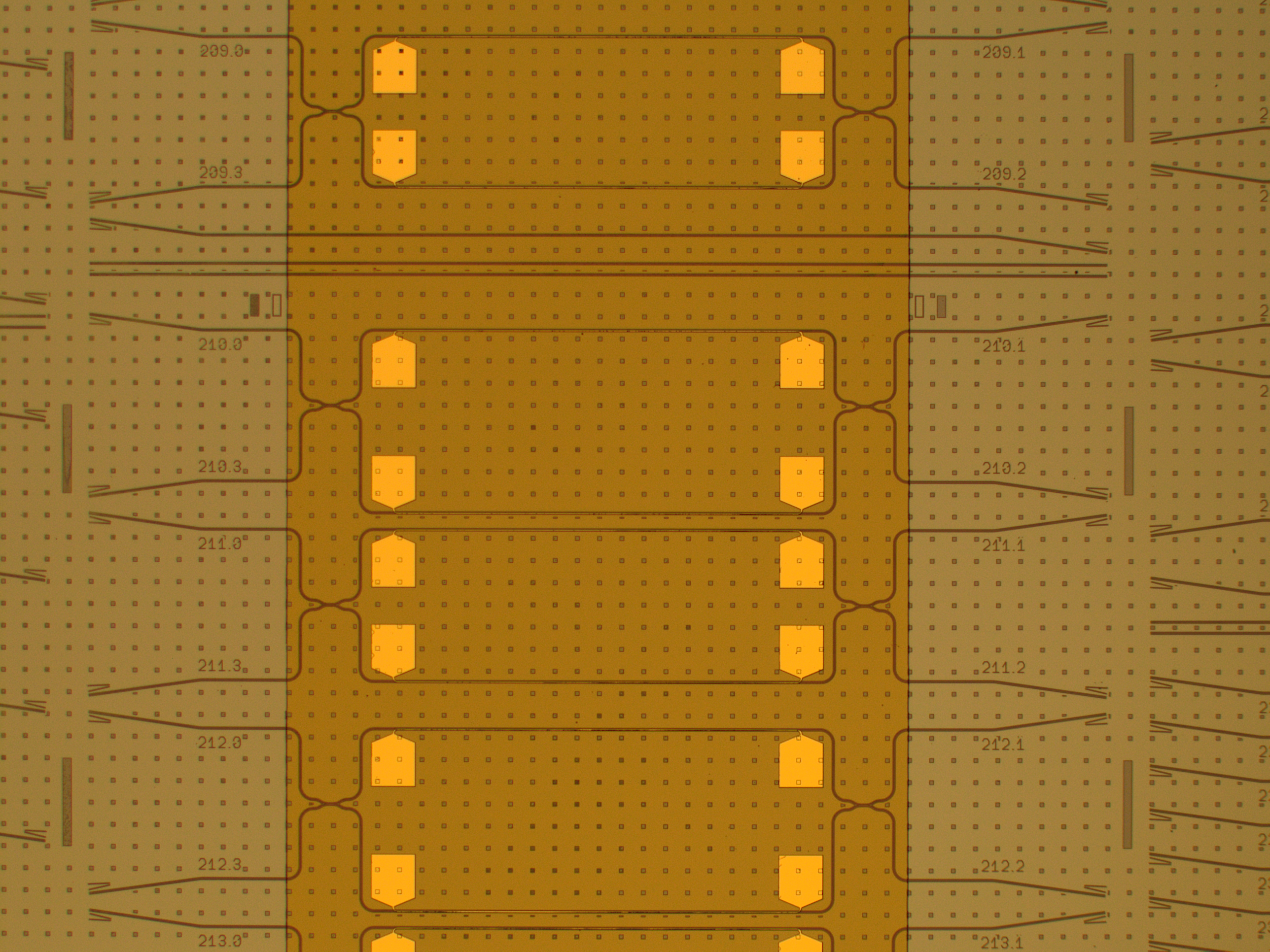}}
\subfigure[\protect  \label{Spectrum_Shift}]{
\includegraphics[width=0.45\textwidth]{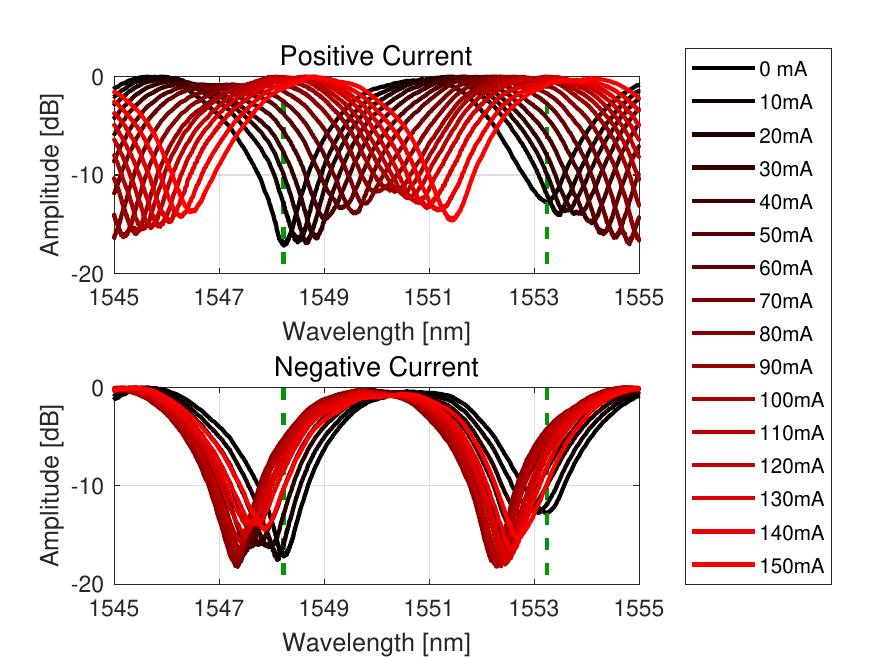}}
\\
\subfigure[\protect  \label{ThermoOptic_and_MagnetoOptic}]{
\includegraphics[width=0.45\textwidth]{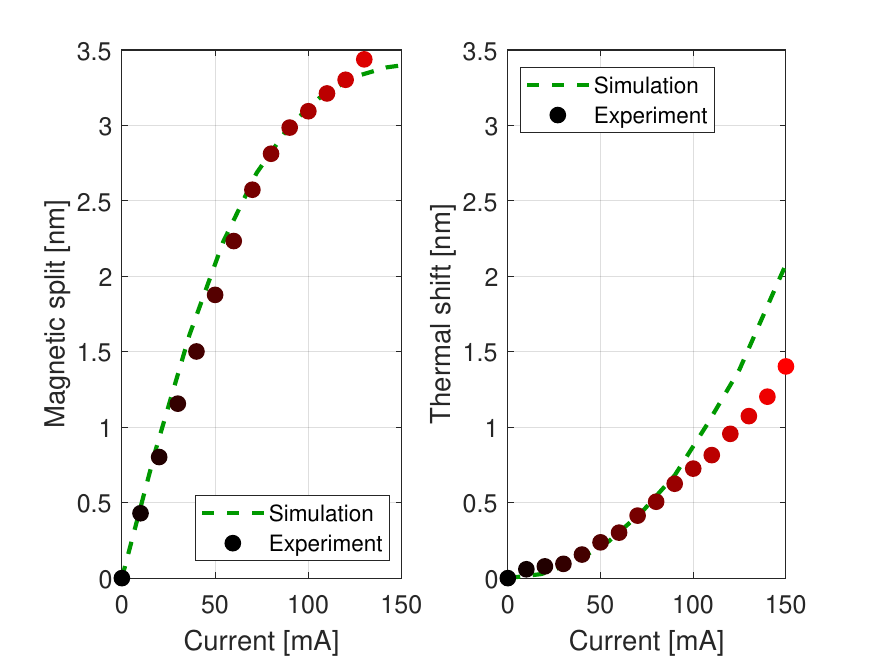}}
\subfigure[\protect  \label{MagneticField_Exp_vs_Sim}]{
\includegraphics[width=0.45\textwidth]{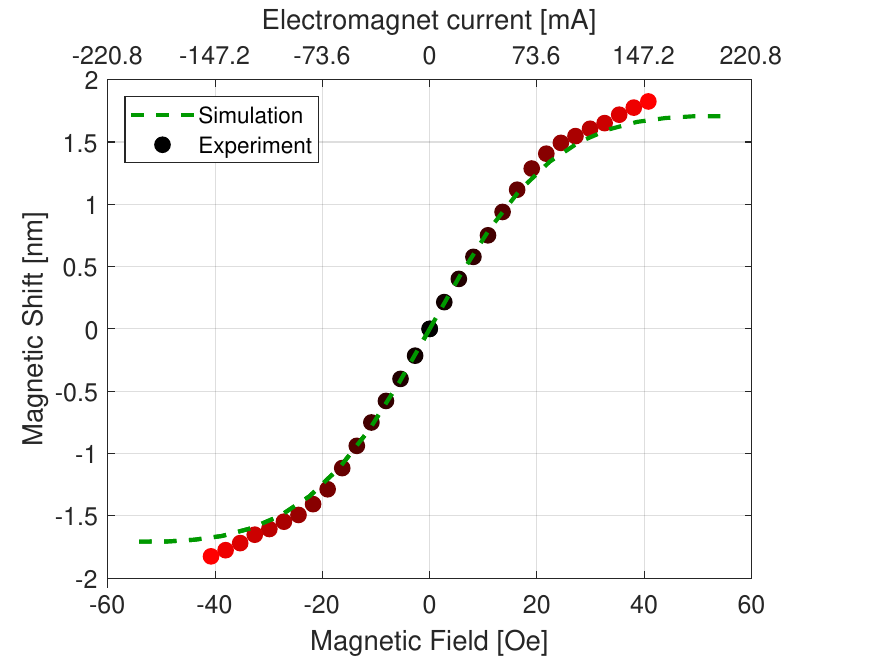}}
\caption{Experimental characterization of the heterogeneous Ce:YIG-on-Si integrated magnetometer. (a)~Fabricated Mach-Zehnder interferometer. (b)~Measured spectral shift induced by current applied to one electromagnet: positive current corresponds to the same direction as light propagation, while negative current is in the opposite direction. (c)~Comparison between simulated and experimental results, showing both the magnetic splitting and thermal shift of the measured spectra. (d)~Magnetic field inferred from the magnetically induced spectral shift as a function of the applied current. In this plot, the thermal contribution has been removed.}\label{Fig:Exp_CeYIH_on_Si}
\end{figure*}

The proposed optical magnetometer is experimentally validated using the MZI shown in Fig.~\ref{Fabricated_MZIs}. In this proof-of-concept device, both arms of a silicon MZI are covered with a 400~nm-thick layer of Ce:YIG, with a 5~$\mu$m-thick top-cladding of SGGG. Two integrated gold electromagnets serve the purpose of generating the in-plane magnetic field. Fabrication process of the device is similar to the one reported in~\cite{huang2016electrically}. In the ultimate magnetometer prototype, the MO material will be bonded solely to one arm of the MZI. In fact, when the MO covers the entire interferometer, the resulting NRPS induced by a homogeneous external magnetic field would be null. However, the device in Fig.~\ref{Fabricated_MZIs} offers the advantage of easy fabrication since no selective etching or precise alignment is required, and the magnetic field can be independently applied to each MZI arm separately.

To have high-quality Ce:YIG, the crystal is grown on a separate substituted gadolinium gallium garnet (SGGG) substrate, and subsequently transferred to the SOI wafer using a flip-chip process based on plasma-activated SiO$_2$–SiO$_2$ covalent bonding at low temperatures~\cite{liang2008low,tien2011silicon}. Although Ce:YIG has been shown to be deposited successfully on silicon waveguides using techniques such as pulsed laser deposition~\cite{bi2011chip, zhang2019monolithic} and sputtering~\cite{block2013growth}, growing it on SGGG yields larger Faraday rotation constants and reduced propagation loss.

To characterize the MZI, light from a tunable continuous-wave single-mode laser is injected through an input port on the left-hand side, and the output power is collected at the right-hand side using an InGaAsP photodetector. The measured spectra are normalized to their maximum values as shown in Fig.\ref{Spectrum_Shift} for a current in the electromagnet ranging from 0 to $\pm$150mA. The spectrum measured at 0~mA is used as a reference case (shown in black in Fig.\ref{Spectrum_Shift}). Small-magnitude electrical currents induce a linear shift in the spectra. The spectra are red-shifted when the magnetic field is positive (resulting from a positive electromagnet current) and blue-shifted when the magnetic field is directed in the opposite direction (due to a negative electromagnet current). When the current amplitude exceeds 50~mA, the local heating produced by the electromagnet becomes more significant. For larger currents, local heating causes the spectra to be (quadratically) red-shifted for both positive and negative electromagnetic currents.


To gauge the magnetic field detectable, we dissect the spectral shift by separating its magnetic and thermal contributions, and then compare these terms against a numerical model. We observed that the spectra measured for two current values with the same amplitude but opposite signs exhibit opposing MO shifts ($\pm|I|\sim\pm|H|\sim\pm|\Delta\lambda_{MO}|$) while sharing the same thermo-optical (TO) shift ($|I|^2\sim \Delta T \sim|\Delta\lambda_{TO}|$).

Following~\cite{pintus2017microring}, the thermal and magnetic shifts are extracted from the spectral minima measured at currents of equal amplitude but opposite direction. The thermal shift is taken as the average of the shifts at positive and negative currents, while the magnetic shift is half the separation between them, called the magnetic split in Fig.~\ref{ThermoOptic_and_MagnetoOptic}.

To reproduce the electromagnet behavior and capture the overall characteristic of the device, a model is constructed within COMSOL (see the supplementary material). From these numerical results, we calculate the corresponding magnetic split and thermal shifts, comparing them with the experimental results illustrated in Fig.\ref{ThermoOptic_and_MagnetoOptic}. We can assess the magnetic shift as a function of the applied magnetic field by isolating the thermal contribution. The experimental setup employs an applied current, and the magnetic field is derived from a simulated relationship with the current. The results shown in Fig.\ref{MagneticField_Exp_vs_Sim} prove the feasibility of measuring the direction and amplitude of a magnetic field within the $\pm$40~Oe ($\pm$4~mT) range.

Based on the minimum detectable field derived in the previous section, we estimate a dynamic range of approximately 80~dB.


\section{Result discussion}

As outlined above, the maximum detectable field is limited by the saturation of the MO material, while the minimum sensitivity is constrained by noise. In this section, we analyze the fundamental limitations of the device and discuss approaches to enhance its sensitivity.

\begin{figure*}[!thb]
\centering
\subfigure[\protect   \label{Sensitivity_ClassicalLimit_Length_and_MOLoss}]{
\includegraphics[width=0.5\textwidth]{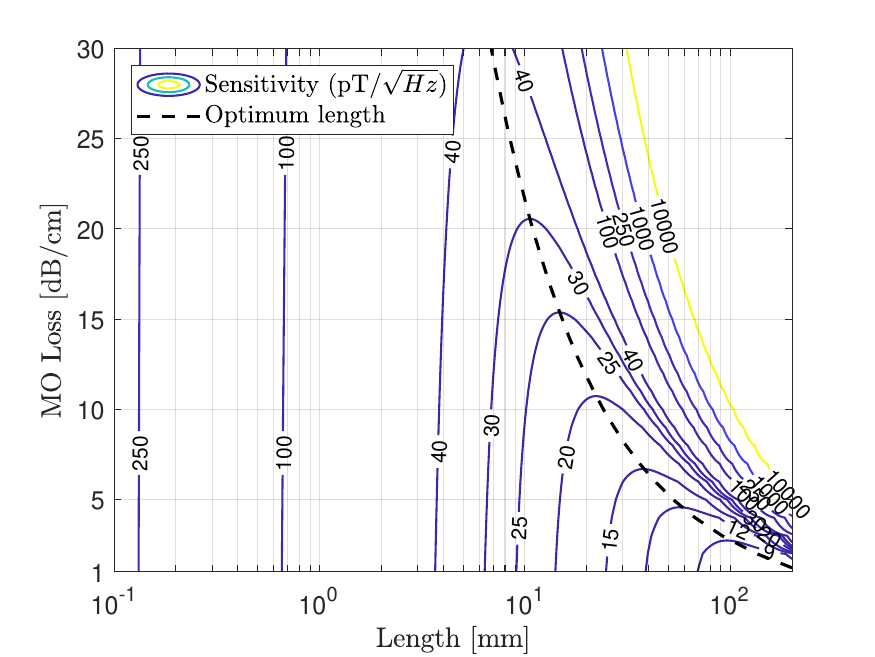}}
\subfigure[\protect  \label{Sensitivity_Limit_vs_Freq}]{
\includegraphics[width=0.5\textwidth]{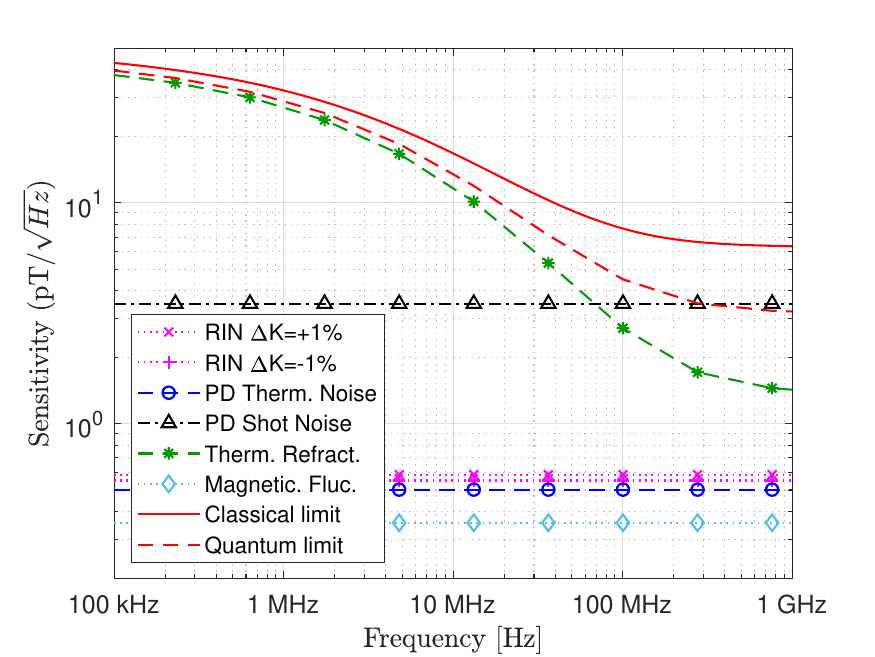}}
\caption{Sensitivity analysis. (a)~Sensitivity as a function of optical loss in the MO material and the arm length of the Mach–Zehnder interferometer. (b)~Classical and quantum limits to the sensitivity of the Ce:YIG-on-Si magnetometer as a function of frequency. Simulations assume a nominal MO arm length of 3 mm with a propagation loss of 25 dB/cm and an input optical power of 20~mW.}\label{Fig:Sensitivity_optimization}
\end{figure*}

The sensitivity depends on both the Faraday rotation and the optical absorption of the MO material. Since the conversion gain 
$G$ is proportional to the Faraday rotation, employing an MO material with larger Faraday rotation lowers the minimum detectable magnetic field (see~\eqref{eq:sensitivity}). The effects of propagation loss is more nuanced. Lower propagation loss allows for a longer arm length 
$L$, reducing thermo-refractive and magnetic fluctuation noise (see Fig.~\ref{Sensitivity_Limit_PAPER}). At the same time, lower propagation loss increases the average power attenuation $A\to1$, leading to reduced shot and thermal noise at the PD. The results of this analysis is shown in Fig.~\ref{Sensitivity_ClassicalLimit_Length_and_MOLoss}, where sensitivity is computed for different arm length and MO propagation loss. 

The results presented in Fig.~\ref{Fig:Sim_CeYIH_on_Si} highlight that the sensitivity is primarily limited by thermo-refractive noise. The power spectral density of this noise contribution is proportional to $k^{-1}(T \cdot dn/dT)^2$, where $k$ is the thermal conductivity, $T$ is the temperature and $dn/dT$ is the thermo-optic coefficient~\cite{wanser1992fundamental,bartolo2012thermal}.

To reduce the thermo-refractive noise, one can consider replacing silicon with silicon nitride (SiN). SiN has a thermo-optic coefficient of $2.45 \cdot 10^{-5}$/K~\cite{arbabi2013measurements}, which is approximately ten times lower than that of silicon. For this platform, we found that the optimal waveguide cross-section that maximizes the phase shift while ensuring single-mode operation is 1.6~$\mu$m wide and 900~nm tall, with a 325~nm thick Ce:YIG layer on top. While replacing Si with SiN significantly reduces the power spectral density of thermo-refractive noise, the trade-off is a lower refractive index waveguide, which diminishes both the optical mode confinement and the MO interaction. This produces a lower conversion gain, $G$, resulting in a smaller overall sensitivity. Additional details are provided in the supplementary material.

An alternative strategy for reducing thermo-refractive noise is lowering the operating temperature. To quantify this effect, we developed a detailed model that simulates the noise power spectral density as a function of temperature, incorporating literature-reported values for thermal conductivity and heat capacity of the constituent materials (see the supplementary materials for more details). Although the thermo-refractive noise decreases substantially at lower temperatures, Ce:YIG shows temperature-dependent magnetic properties, for example Ce:YIG grown on GGG exhibited a change in magnetic anisotropy and hysteretic magnetic behavior below 200~K~\cite{lage2017temperature}. This hysteresis complicates the reliable detection of magnetic fields, thereby limiting the practical benefits of cryogenic cooling.

To reduce noise, we can leverage the minimal size of such an integrated magnetometer. Averaging the detected field over an array of $N_d$ magnetometers reduces the noise by a factor of $\sqrt{N_d}$. This approach is analogous to the improvement in paralleling amplifiers to improve the signal-to-noise ratio~\cite{Scott2015}.

It is also worth noting that the power spectral density of thermo-refractive noise drops significantly at frequencies above 10~MHz, as shown in Fig.~\ref{PhaseNoiseSpectrum}. To take advantage of this behavior, we analyzed the high-frequency regime for an MZI with a 3~mm-long MO arm. As shown in Fig.~\ref{Sensitivity_Limit_vs_Freq}, the sensitivity is approximately 40~pT/$\sqrt{\mathrm{Hz}}$ at low frequencies, improving to about 6~pT/$\sqrt{\mathrm{Hz}}$ at frequency larger than 200~MHz, where it becomes limited by the PD shot noise (i.e., the quantum limit).

To overcome this quantum limit, squeezed light can be injected into the system~\cite{ganapathy2023broadband}. For instance, employing 10~dB of squeezing can enhance sensitivity at high frequencies, allowing detection levels below 3~pT/$\sqrt{\mathrm{Hz}}$.


A unique feature of this magnetometer is its ultra-wide operational bandwidth, which spans from DC up to the ultra-high frequency range, limited only by the ferromagnetic resonance of Ce:YIG ($\sim$2~GHz,~\cite{rombouts2023sub,pintus2022integrated}). This broad bandwidth surpasses that of optomechanical and SQUID magnetometers~\cite{bennett2021precision}, enabling applications such as serving as a magnetic antenna for communications~\cite{pan2017magnecomm,hott2019magnetic,chen2023quantum,hermann2024extending}, measuring the magnetic response from nuclear quadrupole resonance for the identification of explosives and drugs~\cite{garroway1994explosives,czipott1997magnetic,lee2006subfemtotesla,miller2007nuclear}, detecting RF signals in electronic warfare, and locating concealed electronics.

\section{Conclusion}

In this work, we demonstrate that a magnetometer with a magneto-optical cladding can be effectively used to detect external magnetic fields. As a proof of concept, we employ a silicon MZI with a bonded Ce:YIG layer and an integrated electromagnet for magnetic field application. The system operates up to a saturation field of 4~mT and achieves a minimum spectral sensitivity below 40~pT/$\sqrt{\mathrm{Hz}}$. Assuming an integration bandwidth of 1~Hz, this corresponds to a dynamic range of approximately 80~dB.

This magnetometer design enables chip-scale integration with lasers and balanced PDs, and can be seamlessly combined with digital electronics for fast control and signal processing. In a final prototype, the integrated electromagnet could serve as a calibration tool, enabling precise in situ generation of magnetic fields with controllable magnitude and orientation.

A fully packaged device, including a laser and balanced PDs, can be realized within a footprint of just a few square millimeters, with a total weight of only a few grams and laser power consumption below 100~mW~\cite{jones2019heterogeneously}. With such high performance and room-temperature operation, this heterogeneous Ce:YIG-on-silicon magnetometer offers a compelling solution for applications where size, weight, and power constraints are critical.

\begin{backmatter}
\bmsection{Funding}\\
The authors would like to aknowledge the support of the National Science Foundation (NSF) through the QuSeC-TAQS program (2326754), the NSF CAREER program (2045246), the National Aeronautics and Space Administration (NASA) contract 80NSSC21C0258, the Italian Ministerial grant PRIN PNRR 2022 `MAGneto-Optic Integrated Computing - MAGIC' (CUP F53D23008340001), the Fondazione di Sardegna via the `Investigation of Novel Magneto-Optic Materials and Devices for Silicon Photonic Integrated Circuits' project (grant no. CUP F73C23001820007), and the City of Brno through the Junior Research Group Leaders project. Any opinions, findings, conclusions, or recommendations expressed in this material are those of the authors and do not necessarily reflect the views of NASA. Paolo Pintus also acknowledges for support the program `Mobilit\`{a} Giovani Ricercatori' of the University of Cagliari, financed by the Autonomous Region of Sardinia.

\bmsection{Acknowledgments}\\
The authors thank Joe Liang, Stephen Kreger, and Osgar John Ohanian~III of Luna Innovations Incorporated and Riccardo Dettori, Alessio Filippetti, and Claudio Melis of the University of Cagliari for their valuable advice and insightful discussions.

\bmsection{Disclosures}\\
The authors declare that they have no conflicts of interest.

\end{backmatter}

\bibliography{bibliography}

\begin{thebibliography}{10}
\newcommand{\enquote}[1]{``#1''}

\bibitem{canciani2016absolute}
A.~Canciani and J.~Raquet, \enquote{Absolute positioning using the earth's
  magnetic anomaly field,} {\protect\JournalTitle{NAVIGATION: Journal of the
  Institute of Navigation}} \textbf{63}, 111--126 (2016).

\bibitem{claussen2020magnetic}
N.~Claussen, L.~Le, R.~Ashton, K.~Cespedes, A.~Patel, L.~Williams, B.~Miller,
  and J.~Searcy, \enquote{Magnetic navigation for gps-denied airborne
  applications.} Tech. rep., Sandia National Lab.(SNL-NM), Albuquerque, NM
  (United States) (2020).

\bibitem{kitching2018chip}
J.~Kitching, \enquote{Chip-scale atomic devices,}
  {\protect\JournalTitle{Applied Physics Reviews}} \textbf{5}, 031302 (2018).

\bibitem{bennett2021precision}
J.~S. Bennett, B.~E. Vyhnalek, H.~Greenall, E.~M. Bridge, F.~Gotardo,
  S.~Forstner, G.~I. Harris, F.~A. Miranda, and W.~P. Bowen, \enquote{Precision
  magnetometers for aerospace applications: A review,}
  {\protect\JournalTitle{Sensors}} \textbf{21}, 5568 (2021).

\bibitem{fagaly2006superconducting}
R.~Fagaly, \enquote{Superconducting quantum interference device instruments and
  applications,} {\protect\JournalTitle{Review of scientific instruments}}
  \textbf{77}, 101101 (2006).

\bibitem{robbes2006highly}
D.~Robbes, \enquote{Highly sensitive magnetometers—a review,}
  {\protect\JournalTitle{Sensors and Actuators A: Physical}} \textbf{129},
  86--93 (2006).

\bibitem{rondin2014magnetometry}
L.~Rondin, J.-P. Tetienne, T.~Hingant, J.-F. Roch, P.~Maletinsky, and
  V.~Jacques, \enquote{Magnetometry with nitrogen-vacancy defects in diamond,}
  {\protect\JournalTitle{Reports on progress in physics}} \textbf{77}, 056503
  (2014).

\bibitem{allred2002high}
J.~Allred, R.~Lyman, T.~Kornack, and M.~V. Romalis, \enquote{High-sensitivity
  atomic magnetometer unaffected by spin-exchange relaxation,}
  {\protect\JournalTitle{Physical review letters}} \textbf{89}, 130801 (2002).

\bibitem{kim2019cmos}
D.~Kim, M.~I. Ibrahim, C.~Foy, M.~E. Trusheim, R.~Han, and D.~R. Englund,
  \enquote{A {CMOS}-integrated quantum sensor based on nitrogen--vacancy
  centres,} {\protect\JournalTitle{Nature Electronics}} \textbf{2}, 284--289
  (2019).

\bibitem{sturner2021integrated}
F.~M. St{\"u}rner, A.~Brenneis, T.~Buck, J.~Kassel, R.~R{\"o}lver, T.~Fuchs,
  A.~Savitsky, D.~Suter, J.~Grimmel, S.~Hengesbach \emph{et~al.},
  \enquote{Integrated and portable magnetometer based on nitrogen-vacancy
  ensembles in diamond,} {\protect\JournalTitle{Advanced Quantum Technologies}}
  \textbf{4}, 2000111 (2021).

\bibitem{korth2016miniature}
H.~Korth, K.~Strohbehn, F.~Tejada, A.~G. Andreou, J.~Kitching, S.~Knappe, S.~J.
  Lehtonen, S.~M. London, and M.~Kafel, \enquote{Miniature atomic scalar
  magnetometer for space based on the rubidium isotope $^{87}${Rb},}
  {\protect\JournalTitle{Journal of Geophysical Research: Space Physics}}
  \textbf{121}, 7870--7880 (2016).

\bibitem{oelsner2022integrated}
G.~Oelsner, R.~IJsselsteijn, T.~Scholtes, A.~Kr{\"u}ger, V.~Schultze,
  G.~Seyffert, G.~Werner, M.~J{\"a}ger, A.~Chwala, and R.~Stolz,
  \enquote{Integrated optically pumped magnetometer for measurements within
  earth’s magnetic field,} {\protect\JournalTitle{Physical Review Applied}}
  \textbf{17}, 024034 (2022).

\bibitem{thomson2016roadmap}
D.~Thomson, A.~Zilkie, J.~E. Bowers, T.~Komljenovic, G.~T. Reed, L.~Vivien,
  D.~Marris-Morini, E.~Cassan, L.~Virot, J.-M. F{\'e}d{\'e}li \emph{et~al.},
  \enquote{Roadmap on silicon photonics,} {\protect\JournalTitle{Journal of
  Optics}} \textbf{18}, 073003 (2016).

\bibitem{zhou2023prospects}
Z.~Zhou, X.~Ou, Y.~Fang, E.~Alkhazraji, R.~Xu, Y.~Wan, and J.~E. Bowers,
  \enquote{Prospects and applications of on-chip lasers,}
  {\protect\JournalTitle{Elight}} \textbf{3}, 1--25 (2023).

\bibitem{moody20222022}
G.~Moody, V.~J. Sorger, D.~J. Blumenthal, P.~W. Juodawlkis, W.~Loh,
  C.~Sorace-Agaskar, A.~E. Jones, K.~C. Balram, J.~C. Matthews, A.~Laing
  \emph{et~al.}, \enquote{2022 roadmap on integrated quantum photonics,}
  {\protect\JournalTitle{Journal of Physics: Photonics}} \textbf{4}, 012501
  (2022).

\bibitem{moody2020chip}
G.~Moody, L.~Chang, T.~J. Steiner, and J.~E. Bowers, \enquote{Chip-scale
  nonlinear photonics for quantum light generation,} {\protect\JournalTitle{AVS
  Quantum Science}} \textbf{2}, 041702 (2020).

\bibitem{ganapathy2023broadband}
D.~Ganapathy, W.~Jia, M.~Nakano, V.~Xu, N.~Aritomi, T.~Cullen, N.~Kijbunchoo,
  S.~Dwyer, A.~Mullavey, L.~McCuller \emph{et~al.}, \enquote{Broadband quantum
  enhancement of the ligo detectors with frequency-dependent squeezing,}
  {\protect\JournalTitle{Physical Review X}} \textbf{13}, 041021 (2023).

\bibitem{aasi2015advanced}
B.~Abbott, M.~Abernathy, R.~Adhikari \emph{et~al.}, \enquote{Advanced ligo,}
  {\protect\JournalTitle{Classical and quantum gravity}} \textbf{32}, 074001
  (2015).

\bibitem{dotsch2005applications}
H.~D{\"o}tsch, N.~Bahlmann, O.~Zhuromskyy, M.~Hammer, L.~Wilkens, R.~Gerhardt,
  P.~Hertel, and A.~F. Popkov, \enquote{Applications of magneto-optical
  waveguides in integrated optics,} {\protect\JournalTitle{JOSA B}}
  \textbf{22}, 240--253 (2005).

\bibitem{landau2013electrodynamics}
L.~D. Landau, L.~Pitaevskii, and E.~Lifshitz, \emph{Electrodynamics of
  continuous media}, vol.~8 (Elsevier, 2013).

\bibitem{tien2011silicon}
M.-C. Tien, T.~Mizumoto, P.~Pintus, H.~Kromer, and J.~E. Bowers,
  \enquote{Silicon ring isolators with bonded nonreciprocal magneto-optic
  garnets,} {\protect\JournalTitle{Optics express}} \textbf{19}, 11740--11745
  (2011).

\bibitem{bi2011chip}
L.~Bi, J.~Hu, P.~Jiang, D.~H. Kim, G.~F. Dionne, L.~C. Kimerling, and C.~Ross,
  \enquote{On-chip optical isolation in monolithically integrated
  non-reciprocal optical resonators,} {\protect\JournalTitle{Nature Photonics}}
  \textbf{5}, 758--762 (2011).

\bibitem{huang2016electrically}
D.~Huang, P.~Pintus, C.~Zhang, Y.~Shoji, T.~Mizumoto, and J.~E. Bowers,
  \enquote{Electrically driven and thermally tunable integrated optical
  isolators for silicon photonics,} {\protect\JournalTitle{IEEE Journal of
  Selected Topics in Quantum Electronics}} \textbf{22}, 271--278 (2016).

\bibitem{pintus2017microring}
P.~Pintus, D.~Huang, C.~Zhang, Y.~Shoji, T.~Mizumoto, and J.~E. Bowers,
  \enquote{Microring-based optical isolator and circulator with integrated
  electromagnet for silicon photonics,} {\protect\JournalTitle{Journal of
  Lightwave Technology}} \textbf{35}, 1429--1437 (2017).

\bibitem{huang2017dynamically}
D.~Huang, P.~Pintus, C.~Zhang, P.~Morton, Y.~Shoji, T.~Mizumoto, and J.~E.
  Bowers, \enquote{Dynamically reconfigurable integrated optical circulators,}
  {\protect\JournalTitle{Optica}} \textbf{4}, 23--30 (2017).

\bibitem{zhang2019monolithic}
Y.~Zhang, Q.~Du, C.~Wang, T.~Fakhrul, S.~Liu, L.~Deng, D.~Huang, P.~Pintus,
  J.~Bowers, C.~A. Ross \emph{et~al.}, \enquote{Monolithic integration of
  broadband optical isolators for polarization-diverse silicon photonics,}
  {\protect\JournalTitle{Optica}} \textbf{6}, 473--478 (2019).

\bibitem{pintus2019broadband}
P.~Pintus, D.~Huang, P.~A. Morton, Y.~Shoji, T.~Mizumoto, and J.~E. Bowers,
  \enquote{Broadband te optical isolators and circulators in silicon photonics
  through ce: Yig bonding,} {\protect\JournalTitle{Journal of Lightwave
  Technology}} \textbf{37}, 1463--1473 (2019).

\bibitem{pintus2022integrated}
P.~Pintus, L.~Ranzani, S.~Pinna, D.~Huang, M.~V. Gustafsson, F.~Karinou, G.~A.
  Casula, Y.~Shoji, Y.~Takamura, T.~Mizumoto, and J.~E. Bowers, \enquote{An
  integrated magneto-optic modulator for cryogenic applications,}
  {\protect\JournalTitle{Nature Electronics}} \textbf{5}, 604--610 (2022).

\bibitem{rombouts2023sub}
M.~P. Rombouts, F.~Karinou, P.~Pintus, D.~Huang, J.~E. Bowers, and
  N.~Calabretta, \enquote{A sub-picojoule per bit integrated magneto-optic
  modulator on silicon: Modeling and experimental demonstration,}
  {\protect\JournalTitle{Laser \& Photonics Reviews}} \textbf{17}, 2200799
  (2023).

\bibitem{yajima2023high}
S.~Yajima, N.~Nishiyama, and Y.~Shoji, \enquote{High-speed modulation in a
  waveguide magneto-optical switch with impedance-matching electrode,}
  {\protect\JournalTitle{Optics Express}} \textbf{31}, 16243--16250 (2023).

\bibitem{murai2020nonvolatile}
T.~Murai, Y.~Shoji, N.~Nishiyama, and T.~Mizumoto, \enquote{Nonvolatile
  magneto-optical switches integrated with a magnet stripe array,}
  {\protect\JournalTitle{Optics Express}} \textbf{28}, 31675--31685 (2020).

\bibitem{pintus2025integrated}
P.~Pintus, M.~Dumont, V.~Shah, T.~Murai, Y.~Shoji, D.~Huang, G.~Moody, J.~E.
  Bowers, and N.~Youngblood, \enquote{Integrated non-reciprocal magneto-optics
  with ultra-high endurance for photonic in-memory computing,}
  {\protect\JournalTitle{Nature Photonics}} \textbf{19}, 54--62 (2025).

\bibitem{wanser1992fundamental}
K.~H. Wanser, \enquote{Fundamental phase noise limit in optical fibres due to
  temperature fluctuations,} {\protect\JournalTitle{Electronics letters}}
  \textbf{28}, 53 (1992).

\bibitem{bartolo2012thermal}
R.~E. Bartolo, A.~B. Tveten, and A.~Dandridge, \enquote{Thermal phase noise
  measurements in optical fiber interferometers,} {\protect\JournalTitle{IEEE
  Journal of Quantum Electronics}} \textbf{48}, 720--727 (2012).

\bibitem{callen1951irreversibility}
H.~B. Callen and T.~A. Welton, \enquote{Irreversibility and generalized noise,}
  {\protect\JournalTitle{Physical Review}} \textbf{83}, 34 (1951).

\bibitem{de2013non}
S.~R. De~Groot and P.~Mazur, \emph{Non-equilibrium thermodynamics} (Courier
  Corporation, 2013).

\bibitem{levin2008fluctuation}
Y.~Levin, \enquote{Fluctuation--dissipation theorem for thermo-refractive
  noise,} {\protect\JournalTitle{Physics Letters A}} \textbf{372}, 1941--1944
  (2008).

\bibitem{landau2013statistical}
L.~D. Landau and E.~M. Lifshitz, \emph{Statistical Physics: Volume 5}, vol.~5
  (Elsevier, 2013).

\bibitem{brown1963thermal}
W.~F. Brown~Jr, \enquote{Thermal fluctuations of a single-domain particle,}
  {\protect\JournalTitle{Physical review}} \textbf{130}, 1677 (1963).

\bibitem{leliaert2017adaptively}
J.~Leliaert, J.~Mulkers, J.~De~Clercq, A.~Coene, M.~Dvornik, and
  B.~Van~Waeyenberge, \enquote{Adaptively time stepping the stochastic
  landau-lifshitz-gilbert equation at nonzero temperature: Implementation and
  validation in {MuMax3},} {\protect\JournalTitle{Aip Advances}} \textbf{7},
  125010 (2017).

\bibitem{abo2013definition}
G.~S. Abo, Y.-K. Hong, J.~Park, J.~Lee, W.~Lee, and B.-C. Choi,
  \enquote{Definition of magnetic exchange length,} {\protect\JournalTitle{IEEE
  Transactions on Magnetics}} \textbf{49}, 4937--4939 (2013).

\bibitem{srinivasan2014integrated}
S.~Srinivasan and J.~E. Bowers, \enquote{Integrated high sensitivity hybrid
  silicon magnetometer,} {\protect\JournalTitle{IEEE Photonics Technology
  Letters}} \textbf{26}, 1321--1324 (2014).

\bibitem{srinivasan2014design}
S.~Srinivasan, R.~Moreira, D.~Blumenthal, and J.~E. Bowers, \enquote{Design of
  integrated hybrid silicon waveguide optical gyroscope,}
  {\protect\JournalTitle{Optics express}} \textbf{22}, 24988--24993 (2014).

\bibitem{huang2015compact}
D.~Huang, S.~Srinivasan, and J.~E. Bowers, \enquote{Compact tb doped fiber
  optic current sensor with high sensitivity,} {\protect\JournalTitle{Optics
  express}} \textbf{23}, 29993--29999 (2015).

\bibitem{gundavarapu2017interferometric}
S.~Gundavarapu, M.~Belt, T.~A. Huffman, M.~A. Tran, T.~Komljenovic, J.~E.
  Bowers, and D.~J. Blumenthal, \enquote{Interferometric optical gyroscope
  based on an integrated si3n4 low-loss waveguide coil,}
  {\protect\JournalTitle{Journal of Lightwave Technology}} \textbf{36},
  1185--1191 (2017).

\bibitem{pintus2013integrated}
P.~Pintus, F.~Di~Pasquale, and J.~E. Bowers, \enquote{Integrated te and tm
  optical circulators on ultra-low-loss silicon nitride platform,}
  {\protect\JournalTitle{Optics express}} \textbf{21}, 5041--5052 (2013).

\bibitem{pintus2011design}
P.~Pintus, M.-C. Tien, and J.~E. Bowers, \enquote{Design of magneto-optical
  ring isolator on soi based on the finite-element method,}
  {\protect\JournalTitle{IEEE Photonics Technology Letters}} \textbf{23},
  1670--1672 (2011).

\bibitem{pintus2014accurate}
P.~Pintus, \enquote{Accurate vectorial finite element mode solver for
  magneto-optic and anisotropic waveguides,} {\protect\JournalTitle{Optics
  express}} \textbf{22}, 15737--15756 (2014).

\bibitem{komma2012thermo}
J.~Komma, C.~Schwarz, G.~Hofmann, D.~Heinert, and R.~Nawrodt,
  \enquote{Thermo-optic coefficient of silicon at 1550 nm and cryogenic
  temperatures,} {\protect\JournalTitle{Applied Physics Letters}} \textbf{101},
  041905 (2012).

\bibitem{elshaari2016thermo}
A.~W. Elshaari, I.~E. Zadeh, K.~D. J{\"o}ns, and V.~Zwiller,
  \enquote{Thermo-optic characterization of silicon nitride resonators for
  cryogenic photonic circuits,} {\protect\JournalTitle{IEEE Photonics Journal}}
  \textbf{8}, 1--9 (2016).

\bibitem{liang2008low}
D.~Liang, A.~W. Fang, H.~Park, T.~E. Reynolds, K.~Warner, D.~C. Oakley, and
  J.~E. Bowers, \enquote{Low-temperature, strong sio 2-sio 2 covalent wafer
  bonding for iii--v compound semiconductors-to-silicon photonic integrated
  circuits,} {\protect\JournalTitle{Journal of Electronic Materials}}
  \textbf{37}, 1552--1559 (2008).

\bibitem{block2013growth}
A.~D. Block, P.~Dulal, B.~J. Stadler, and N.~C. Seaton, \enquote{Growth
  parameters of fully crystallized yig, bi: Yig, and ce: Yig films with high
  faraday rotations,} {\protect\JournalTitle{IEEE Photonics Journal}}
  \textbf{6}, 1--8 (2013).

\bibitem{arbabi2013measurements}
A.~Arbabi and L.~L. Goddard, \enquote{Measurements of the refractive indices
  and thermo-optic coefficients of {Si}$_3${N}$_4$ and {SiO}$_x$ using
  microring resonances,} {\protect\JournalTitle{Optics letters}} \textbf{38},
  3878--3881 (2013).

\bibitem{lage2017temperature}
E.~Lage, L.~Beran, A.~U. Quindeau, L.~Ohnoutek, M.~Kucera, R.~Antos, S.~R.
  Sani, G.~F. Dionne, M.~Veis, and C.~A. Ross, \enquote{Temperature-dependent
  faraday rotation and magnetization reorientation in cerium-substituted
  yttrium iron garnet thin films,} {\protect\JournalTitle{APL Materials}}
  \textbf{5}, 036104 (2017).

\bibitem{Scott2015}
K.~Scott, \enquote{Paralleling amplifiers improves signal-to-noise
  performance,}
  \url{https://www.analog.com/en/resources/technical-articles/paralleling-amplifiers-improves-signal-to-noise-performance.html}
  (2015). Accessed: 2025-08-19.

\bibitem{pan2017magnecomm}
H.~Pan, Y.-C. Chen, G.~Xue, and X.~Ji, \enquote{Magnecomm: Magnetometer-based
  near-field communication,} in \emph{Proceedings of the 23rd Annual
  International Conference on Mobile Computing and Networking,}  (2017), pp.
  167--179.

\bibitem{hott2019magnetic}
M.~Hott, P.~A. Hoeher, and S.~F. Reinecke, \enquote{Magnetic communication
  using high-sensitivity magnetic field detectors,}
  {\protect\JournalTitle{Sensors}} \textbf{19}, 3415 (2019).

\bibitem{chen2023quantum}
X.-D. Chen, E.-H. Wang, L.-K. Shan, S.-C. Zhang, C.~Feng, Y.~Zheng, Y.~Dong,
  G.-C. Guo, and F.-W. Sun, \enquote{Quantum enhanced radio detection and
  ranging with solid spins,} {\protect\JournalTitle{Nature Communications}}
  \textbf{14}, 1288 (2023).

\bibitem{hermann2024extending}
J.~C. Hermann, R.~Rizzato, F.~Bruckmaier, R.~D. Allert, A.~Blank, and D.~B.
  Bucher, \enquote{Extending radiowave frequency detection range with dressed
  states of solid-state spin ensembles,} {\protect\JournalTitle{npj Quantum
  Information}} \textbf{10}, 103 (2024).

\bibitem{garroway1994explosives}
A.~N. Garroway, M.~L. Buess, J.~P. Yesinowski, J.~B. Miller, and R.~A. Krauss,
  \enquote{Explosives detection by nuclear quadrupole resonance (nqr),} in
  \emph{Cargo Inspection Technologies,}  vol. 2276 (SPIE, 1994), pp. 139--149.

\bibitem{czipott1997magnetic}
P.~V. Czipott and M.~D. Iwanowski, \enquote{Magnetic sensor technology for
  detecting mines, uxo, and other concealed security threats,} in
  \emph{Terrorism and Counter-Terrorism Methods and Technologies,}  vol. 2933
  (SPIE, 1997), pp. 67--76.

\bibitem{lee2006subfemtotesla}
S.-K. Lee, K.~Sauer, S.~Seltzer, O.~Alem, and M.~Romalis,
  \enquote{Subfemtotesla radio-frequency atomic magnetometer for detection of
  nuclear quadrupole resonance,} {\protect\JournalTitle{Applied Physics
  Letters}} \textbf{89}, 214106 (2006).

\bibitem{miller2007nuclear}
J.~B. Miller, \enquote{Nuclear quadrupole resonance detection of explosives,}
  in \emph{Counterterrorist detection techniques of explosives,}  (Elsevier,
  2007), pp. 157--198.

\bibitem{jones2019heterogeneously}
R.~Jones, P.~Doussiere, J.~B. Driscoll, W.~Lin, H.~Yu, Y.~Akulova,
  T.~Komljenovic, and J.~E. Bowers, \enquote{Heterogeneously integrated
  {InP}$\backslash$/silicon photonics: fabricating fully functional
  transceivers,} {\protect\JournalTitle{IEEE Nanotechnology Magazine}}
  \textbf{13}, 17--26 (2019).

\end{thebibliography}

\end{document}